\documentstyle[12pt]{article}

\begin{document}

\newcommand{\msqav}{\mbox{$\langle m_{\tilde{q}} \rangle$}}
\newcommand{\ee}{e^+e^-}
\newcommand{\s}{\\ \vspace*{-4mm}}
\newcommand{\non}{\nonumber}
\newcommand{\nn}{\noindent}
\newcommand{\ra}{\rightarrow}
\newcommand{\msq}{\mbox{$m_{\tilde{q}}$}}
\newcommand{\msi}{\mbox{$m_{\tilde{q}_i}$}}
\newcommand{\msj}{\mbox{$m_{\tilde{q}_j}$}}
\newcommand{\mso}{\mbox{$m_{\tilde{q}_1}$}}
\newcommand{\mst}{\mbox{$m_{\tilde{q}_2}$}}
\newcommand{\mg}{\mbox{$M_{\tilde{g}}$}}
\newcommand{\Li}{\mbox{{\rm Li}$_2$}}
\newcommand{\is}{2I_q^{3L}}
\newcommand{\heta}{\tilde{\theta}_q}
\newcommand{\ct}{\cos^2\theta}
\newcommand{\st}{\sin^2\theta}
\renewcommand{\thefootnote}{\fnsymbol{footnote} }

\begin{flushright}
PM 94/31 \\
UdeM-GPP--94--13 \\
December 1994 \\
\end{flushright}

\vspace*{1cm}
\begin{center}
{\large{\bf QCD corrections to scalar quark pair
\\ \vspace*{3mm} production in e$^+$e$^-$ annihilation}}

\vspace*{1cm}

A. ARHRIB\footnote{Also at Universit\'e Cadi Ayyad, Facult\'e des
Sciences--Semlalia, L.P.T.N, B.P. S15 Marrakech, Maroc.}, M.
CAPDEQUI--PEYRANERE

\vspace*{0.3cm}

Physique Math\'ematique et Th\'eorique, U.R.A. 768 du CNRS, \\
Universit\'e de Montpellier II, 34095 Montpellier Cedex 5, France

\vspace*{0.3cm}

and

\vspace*{0.3cm}

A.~DJOUADI

\vspace*{0.3cm}

Groupe de Physique des Particules, Universit\'e de Montr\'eal, \\
Case 6128A, H3C 3J7 Montr\'eal PQ, Canada

\end{center}

\vspace*{1cm}

\begin{abstract}

\vspace*{0.5cm}

\nn We calculate the QCD radiative corrections to the production of the
supersymmetric scalar partners of quarks in $\ee$ annihilation. We include
both the standard gluonic corrections and the genuine supersymmetric QCD
corrections due to quark--gluino loops, and allow for mixing between left--
and right--handed scalar quarks which leads to the possibility that the two
final state particles have different masses. The corrections are found to be
much larger than the ones affecting the production of spin $\frac{1}{2}$
particles.

\end{abstract}
\newpage

\pagestyle{plain}
\renewcommand{\thefootnote}{\arabic{footnote} }
\setcounter{footnote}{0}

\subsection*{1.~Introduction}

One of the best motivated extensions of the Standard Model (SM) of the strong,
electromagnetic and weak interactions is Supersymmetry (SUSY); for reviews see
Refs.~\cite{S1,S2}. Supersymmetric theories provide an elegant way to stabilize
\cite{S3} the huge hierarchy between the electroweak symmetry breaking scale
and the Grand Unification or Planck scales against radiative corrections, and
its minimal version, the minimal supersymmetric Standard Model (MSSM), allows
for a consistent unification of the gauge coupling constants, in contrast to
the nonsupersymmetric SM \cite{S4}; moreover it offers a natural solution of
the cosmological Dark Matter problem \cite{S5}. \s

The MSSM predicts the  existence of scalar partners to all known quarks and
leptons. Since SUSY is broken, these particles can have masses much larger than
the masses of their standard partners; however, naturalness arguments suggest
that the scale of SUSY breaking, and hence the masses of the SUSY particles,
should not exceed ${\cal O}$ (1 TeV). So far, the search for SUSY particles at
colliders has not been successful and under some assumptions, one can only set
lower limits of ${\cal O}(100$ GeV) for squarks \cite{CDF} and below $\sim 50$
GeV for sleptons \cite{LEP} on their masses. Higher energy hadron and $\ee$
colliders will be required to sweep the entire mass range, up to ${\cal O}$ (1
TeV), for the SUSY particles.\s

For a precise prediction of the the production rates of these particles,
radiative
corrections [at least those which are expected to be rather large] should be
incorporated. In particular, because the strong coupling constant is large, the
QCD corrections to squark pair production must be included. The QCD corrections
to squark pair production at proton colliders have been derived recently
\cite{DESY}. At $\ee$ colliders, part of the QCD corrections to squark pair
production [the ones due to virtual gluon exchange and real gluon emission in
the case where the squarks have equal mass] can be adapted from a result
obtained long time ago by Schwinger \cite{SCH} in scalar QED; see also
Ref.~\cite{DH}.
However, in the MSSM, this result is not complete for two reasons. First, it is
well known that the SUSY partners of left--handed and right--handed massive
quarks mix \cite{MIX} and the mixing will allow for the two mass eigenstates to
be non--degenerate; therefore, the Schwinger result must be generalized to
allow for the possibility that the final state squarks have different masses.
Second, in SUSY theories, the gluons have also spin $\frac{1}{2}$
partners, the gluinos, which interact strongly with squarks and quarks; the
gluino--quark--squark interaction will induce a new type of QCD corrections
which has also to be included. \s

It is the purpose of this paper to provide the complete result for the QCD
corrections to squark pair production in $\ee$ annihilation, allowing for the
possibility of having two squarks with different masses in the final state, and
including both the standard gluonic corrections and the corrections due to
quark--gluino loops. \s

The paper is organized as follows. In the next section, we will first set the
notation and exhibit the various couplings as well as the tree--level results
which will be relevant to our discussion. In section 3 and 4, we will give the
analytical results for the standard gluonic corrections and for the corrections
due to quark--gluino loops respectively; results for the equal mass case will
also be given explicitely. In section 5, we will discuss the magnitude of these
QCD effects. A short conclusion is given in section 6. For completeness, we
will list in the Appendix the scalar one, two and three--point functions which
appear in our results.

\renewcommand{\theequation}{2.\arabic{equation}}
\setcounter{equation}{0}

\subsection*{2. Notation and tree--level results}

The production of squark pairs in $e^+ e^-$ collisions proceeds through
$s$--channel photon and $Z$ boson exchanges. The interaction of a neutral
gauge boson $V = \gamma, Z$ with squark current eigenstates are described by
the following lagrangian \cite{S1}:
\begin{eqnarray}
{\cal L}_{\tilde{q}\tilde{q}V} = -ie A^\mu \sum_{i=L,R}
e_{q} \tilde{q}^*_i \stackrel{\leftrightarrow}{\partial_\mu} \tilde{q}_i \
-\frac{ie}{s_Wc_W} Z^\mu \sum_{i=L,R} (I^{3L}_q - e_{q}
s_W^2) \tilde{q}^*_i \stackrel{\leftrightarrow}{\partial_\mu} \tilde{q}_i
\end{eqnarray}
\nn with $I_q^{3L} = \pm 1/2$ and $e_q$ the weak isospin and the electric
charge
[in units of the proton charge $e$] of the squark, respectively; $s_W^2=1-
c_W^2=\sin^2\theta_W$. As previously mentioned, the supersymmetric partners of
left-- and right--handed massive quarks will mix \cite{MIX}, the mass
eigenstates $\tilde{q}_1$ and $\tilde{q}_2$ being related to the current
eigenstates $\tilde{q}_L$ and
$\tilde{q}_R$ by
\begin{eqnarray}
\tilde{q}_1=\tilde{q}_L \cos \heta +\tilde{q}_R \sin \heta \ \ , \hspace*{1cm}
\tilde{q}_2=- \tilde{q}_L \sin \heta +\tilde{q}_R \cos \heta
\end{eqnarray}
The mixing angle $\heta$ is proportional to the mass of the quark $q$. In the
case of the supersymmetric partners of the light quarks, mixing between the
current eigenstates can therefore be neglected.
However, mixing between top squarks can be
sizeable and allows one of the two mass eigenstates
to be much lighter than the top quark. Bottom squark
mixing can also be significant if the ratio of the vacuum expectation values of
the two Higgs fields which give separately masses to isospin up and isospin
down type fermions is very large. After the introduction of this squark mixing,
the coupling between a gauge boson $V$ and two squarks $\tilde{q}_i$ and
$\tilde{q}_j$ with $i,j=1,2$ are given by [the directions of the momenta are
shown in Fig.~1a]
\begin{eqnarray}
\Gamma^\mu_{\tilde{q}_i \tilde{q}_j \gamma} = -ie e_q (k_1+k_2)^\mu \ \delta_{
ij} \ \ , \hspace*{0.6cm}
\Gamma^\mu_{\tilde{q}_i \tilde{q}_j Z} = - \frac{ie}{4s_Wc_W} (k_1+k_2)^\mu \
a_{ij}  \non
\end{eqnarray}
\begin{eqnarray}
a_{11}= 2(2I_q^{3L} \cos^2 \heta -2s_W^2 e_q) \ , \
a_{22}= 2(2I_q^{3L} \sin^2 \heta -2s_W^2 e_q) \ , \
a_{12}= a_{21}= -2 I_q^{3L} \sin 2\heta
\end{eqnarray}
In our convention, the couplings of quarks to the photon and the $Z$ boson are
given by
\begin{eqnarray}
\Gamma^\mu_{q \bar{q} \gamma} = -ie e_q \gamma^\mu \ \ ,  \hspace*{0.6cm}
\Gamma^\mu_{q\bar{q} Z} &=& -\frac{ie}{4s_Wc_W} \gamma^\mu( v_q - a_q \gamma_5)
\nonumber \\
v_q= 2I_q^{3L} -4 s_W^2 e_q \ \ & , & \ \  a_q= 2I_q^{3L}
\end{eqnarray}

\vspace*{2mm}

\nn When discussing QCD corrections we will also need the squark--squark--gluon
and squark--squark--gluon--electroweak gauge boson interaction lagrangians
which read \cite{S1}
\begin{eqnarray}
{\cal L}_{g\tilde{q}\tilde{q}} &=& -i g_S T^a \
g_a^\mu [ \tilde{q}_1^* \stackrel{\leftrightarrow}{\partial_\mu} \tilde{q}_1
    + \tilde{q}_2^* \stackrel{\leftrightarrow}{\partial_\mu} \tilde{q}_2 ]
\nonumber \\
{\cal L}_{\tilde{q_i}\tilde{q_i}gV} &=& 2g_S e T^a g_a^\mu \sum_{i=L,R} \left[
e_q A_\mu \tilde{q}_i^* \tilde{q}_i + \frac{1}{c_Ws_W} (I_q^{3i}-e_q s_W^2)
Z_\mu \tilde{q}_i^* \tilde{q}_i \right]
\end{eqnarray}
Finally, we need the squark--quark--gluino interaction lagrangian which, in the
presence of squark mixing, is given by \cite{HK}
\begin{eqnarray}
{\cal L}_{\tilde{g}\tilde{q}q} =  -i\sqrt{2}g_S T^a \overline{q} \ \left[
(\tilde{v}_1 + \tilde{a}_1 \gamma_5) \tilde{q_1} +
(\tilde{v}_2 + \tilde{a}_2 \gamma_5) \tilde{q_2} \ \right] \tilde{g}^a
+ {\rm h.c.}
\end{eqnarray}
\nn where $g_S$ is the strong coupling constant, $T^a$ are SU(3)$_C$
generators and $\tilde{v}_i, \tilde{a}_i$ are given by
\begin{eqnarray}
\tilde{v}_1 = \frac{1}{2} (\cos \heta -\sin \heta) = \tilde{a}_2 \ \ \ ,  \ \ \
\tilde{a}_1 = \frac{1}{2} (\cos \heta +\sin \heta) = -\tilde{v}_2 \ \ \
\end{eqnarray}

\nn In the Born approximation, the total cross section for the production of a
pair of squarks [possibly with different masses when produced through
$s$--channel Z boson exchange] $\tilde{q}_i$ and $\tilde{q}_j$ in $\ee$
annihilation, including the finite width of the $Z$ boson, is given by
\begin{eqnarray}
\sigma^B (e^+ e^- \rightarrow \tilde{q}_i \tilde{q}_j) = \frac{\pi \alpha^2}{s}
\lambda^{3/2}_{ij} \left[ e_q^2 \delta_{ij} -\frac{e_q v_e a_{ij}\delta_{ij}}
{16c_W^2s_W^2}\frac{s}{s-M_Z^2} + \frac{(a_e^2+v_e^2)a_{ij}^2}{256s_W^4c_W^4}
\frac{s^2}{(s-M_Z^2)^2+ \Gamma^2_Z M_Z^2} \right]
\end{eqnarray}
with $s= q^2=(k_1-k_2)^2$ the center of mass energy of the $e^+ e^-$
collider and $\lambda_{ij}$ the usual two--body phase--space function,
\begin{eqnarray}
\lambda_{ij}= (1-\mu_i^2 - \mu_j^2)^2 - 4 \mu_i^2 \mu_j^2
\ \ \ , \ \ \mu^2_{i,j}= \tilde{m}^2_{q_{i,j}}/s
\end{eqnarray}
In the case where the mixing between squarks is neglected [as is the case for
the supersymmetric partners of light quarks and for stop and sbottom quarks
if the parameter which describes the strength of the non--SUSY trilinear scalar
interactions $A_{t,b}$ and the SUSY Higgs(ino) mass which also enters trilinear
scalar vertices $\mu$, are set to zero]
the total cross section for left-- and right--handed squark pair production
simplifies to \cite{DH,HK}
\begin{eqnarray}
\sigma^B( e^+ e^- \rightarrow \tilde{q_i} \tilde{q_i}) = \frac{\pi \alpha^2}{s}
\beta^3_q \left[ e_q^2 - \frac{e_q v_e v_{q_i} }{16c_W^2s_W^2}\frac{s}{s-M_Z^2}
+ \frac{(a_e^2+v_e^2)v_{q_i}^2}{256s_W^4c_W^4}\frac{s^2}{(s-M_Z^2)^2+\Gamma^2_Z
M_Z^2} \right]
\end{eqnarray}
with
\begin{eqnarray}
v_{\tilde{q}_1}= v_{\tilde{q}_L} = v_q+ a_q \ \ \ , \ \ \ \
v_{\tilde{q}_2}= v_{\tilde{q}_R} = v_q- a_q
\end{eqnarray}
Note that, as it is expected for the production of scalar particles in $\ee$
annihilation, the cross section is proportional to the third power of the
velocity of the final squarks, $\beta_q= (1-4\tilde{m}_q^2/s)^{1/2}$, and
therefore is strongly suppressed near threshold.
The angular distribution is also typical for spin--zero particle production,
it is given by [$\theta$ is the scattering angle]
\begin{eqnarray}
\frac{{\rm d} \sigma^B}{ {\rm d} \cos \theta}(\ee \ra \tilde{q}_i \tilde{q}_j)
= \frac{3}{4} \sin^2 \theta \ \sigma^B (\ee \ra \tilde{q}_i \tilde{q}_j)
\end{eqnarray}

\vspace*{0.2cm}

\nn In order to include QCD corrections at first order in $\alpha_S$, one
needs to
consider the diagrams in Fig.~1b--1e, the contributions of which will be
discussed in the next two sections.

\renewcommand{\theequation}{3.\arabic{equation}}
\setcounter{equation}{0}

\subsection*{3. Gluonic corrections}

The first set of ${\cal O}(\alpha_S)$ contributions is due to the standard
QCD gluonic corrections. They consist first on the interference between the
contribution of the tree--level diagram Fig.~1a and the sum of the vertex
correction Fig.~1b and the squark wave function renormalization Fig.~1c, where
a gluon is exchanged in the squark legs. These two contributions are separately
ultraviolet divergent and these divergences are regulated using the dimensional
regularization scheme. The sum of the two contributions is, as it should be,
ultraviolet finite but it is infrared divergent; this last divergence is
regulated by introducing a fictitious mass $m_g$ for the gluon. One has also to
include the contribution of the bremsstrahlung diagrams where a gluon is
emitted from the external squark lines, Fig.~1d. This contribution is also
infrared divergent, but the sum of the virtual and real corrections is infrared
finite as expected. Summing all corrections, the cross section at ${\cal O}
(\alpha_S)$ can be written as
\begin{eqnarray}
\sigma (e^+ e^- \rightarrow \tilde{q}_i \tilde{q}_j) =
\sigma^B (e^+ e^- \rightarrow \tilde{q}_i \tilde{q}_j) \left[1 +\frac{4}{3}
\frac{\alpha_S}{\pi} \Delta_{ij} \right] \ \ \ , \ \ \Delta_{ij}= \Delta_{ij}^V
 +\Delta_{ij}^R
\end{eqnarray}
where $\Delta^V_{ij}$ corresponds to the contribution of the virtual exchange
of
gluons and $\Delta^R_{ij}$ to the contribution of gluon emission from the final
state squarks.

\bigskip

\nn {\bf 3.1 Virtual Corrections}

\bigskip

\nn Summing the vertex correction and the squark wave function renormalization
contributions, which as previously mentioned are separately ultraviolet
divergent, one obtains for the interference between the tree level diagram and
the ones of Fig.~1b--1c [after normalizing to the Born term and factorizing out
$4\alpha_S/3\pi$]
\begin{eqnarray}
\Delta_{ij}^V &=& \log \frac{\mu_i \mu_j }{\mu_g^2} -2+\frac{1}{\lambda_{ij}^
{1/2} } (1- \mu_i^2-\mu_j^2) \log \frac{1-\mu_i^2 -\mu_j^2+\lambda_{ij}^{1/2}}
{1-\mu_i^2 -\mu_j^2 -\lambda_{ij}^{1/2}}
\nonumber \\
&& + \frac{1}{2 \lambda_{ij}^{1/2}} (1-\mu_i^2-\mu_j^2) \left( \log \frac{
1-\mu_i^2-\mu_j^2 +\lambda_{ij}^{1/2} }{2 \mu_i \mu_j} \log \mu_g^2 +
F_{ij}^V \right)
\end{eqnarray}
with the scaled variable $\mu_g^2= m_g^2/s$ where $m_g$ is a fictitious gluon
mass introduced to regularize the infrared divergence. The ultraviolet and
infrared finite function $F_{ij}^V$ is given by
\begin{eqnarray}
F_{ij}^V &=& \frac{1}{2} [ \log^2(1-y_1) - \log^2(-y_1) -\log^2(1-y_2) +
\log^2(-y_2)] +2 \log \frac{(1-y_1)}{(-y_1)} \log (y_1-y_2) \nonumber \\
&+&  \log (-y_1) \log(-y_2) -\log (1-y_1) \log(1-y_2) + 2 \Li \left(
\frac{y_1}{y_1-y_2} \right) -2\Li \left(\frac{y_1-1}{y_1-y_2} \right)
\end{eqnarray}
with Li$_2$ the usual Spence function defined as Li$_2 (x)= - \int_0^1 dt
\log (1-xt) t^{-1}$ and the variables $y_{1/2}$ given by
\begin{eqnarray}
y_{1/2} = \frac{1}{2} (1+\mu_i^2 -\mu_j^2 \pm \lambda_{ij}^{1/2})
\end{eqnarray}
Note that the term in the second line of eq.~(3.2) [up to a coefficient
$(1-\mu_i^2-\mu_j^2)/s$] is just the Passarino--Veltman \cite{PV,TV} scalar
three--point function in a particular case
\begin{eqnarray}
C_0( k_1^2, k_2^2, s, \tilde{m}_{q_i}, m_g, \tilde{m}_{q_j})=
\frac{1}{2s \lambda_{ij}^{1/2}} \left( 2 \log \frac{1-\mu_i^2-\mu_j^2
+\lambda_{ij}^{1/2} }{2 \mu_i \mu_j } \log \mu_g + F_{ij}^V \right)
\end{eqnarray}
the general form of which is given in the Appendix.

\bigskip

\nn {\bf 3.2 Real Corrections}

\bigskip

\nn In terms of the momenta of the particles in the final state [$k$ is the
four--momentum of the gluon], the amplitude squared of the process $e^+ e^- \ra
\tilde{q}_i \tilde{q}_j g$ is given by
\begin{eqnarray}
|M|^2 (e^+ e^- \rightarrow \tilde{q}_i \tilde{q}_j g) &=& \frac{4 g_S^2
e^4}{3s}
\left[e_q^2 \delta_{ij} - \frac{e_q v_e a_{ij}\delta_{ij}}{16c_W^2 s_W^2}\frac{
s}{s-M_Z^2}+\frac{(a_e^2+v_e^2)a_{ij}^2}{256s_W^4c_W^4} \frac{s^2} {(s-M_Z^2)^2
+ \Gamma^2_Z M_Z^2} \right] \nonumber \\
&\times &\left[ 8- \lambda_{ij} \left( \frac{s^2\mu_i^2}{(k_1 \cdot k)^2}+
\frac{s^2 \mu_j^2}{(k_2 \cdot k)^2}+ \frac{2s}{k_1 \cdot k} +\frac{2s}{k_2
\cdot k} +\frac{s^2(\mu_i^2+
\mu_j^2-1)}{(k_1 \cdot k) (k_2 \cdot k)} \right) \right]
\end{eqnarray}
Note that, as a consequence of gauge invariance, the amplitude for the sum of
the three diagrams of Fig.~1d, should vanish when multiplied by the
four--momentum of the gluon; this provides a good check of the calculation. \s

The integrals over the three--body phase space in the general case where the
masses of the two squarks are unequal [and where both the soft and hard
bremsstrahlung contributions are added up] can be found for instance in
Ref.~\cite{Denner}. Again, after normalizing to the Born term and factorizing
out $4\alpha_S/3\pi$, one obtains for the real corrections
$\Delta_{ij}^R$ with $i \neq j$

\begin{eqnarray}
\Delta_{ij}^R &=& \frac{1-\mu_i^2-\mu_j^2}{\lambda_{ij}^{1/2}} \left[ 2 \log^2
\lambda_0- \log^2 \lambda_1 -\log^2 \lambda_2 + 2\Li(1-\lambda_0^2)
-\Li (1-\lambda_1^2)- \Li (1-\lambda_2^2) \right] \nonumber \\
&+ &  \frac{4}{\lambda_{ij}
^{3/2}} \left[ \frac{1}{4} \lambda_{ij}^{1/2} (1+\mu_i^2+ \mu_j^2) + \mu_i^2
\log \lambda_2+ \mu_j^2 \log \lambda_1+ \mu_i^2 \mu_j^2 \log \lambda_0 \right]
+4 + \frac{1+2\mu_j^2}{\lambda_{ij}^{1/2}} \log \lambda_2  \nonumber \\
&+& \frac{1+2 \mu_i^2}{\lambda_{ij}^{1/2}}  \log \lambda_1+
\left[\frac{(1-\mu_i^2-\mu_j^2)}{\lambda_{ij}^{1/2}} \log\lambda_0 -1
\right] \log \frac{\lambda_{ij}^2}{ \mu_g^2 \mu_i^2 \mu_j^2}
+ \frac{(2+\mu_i^2+\mu_j^2)}{\lambda_{ij}^{1/2}} \log \lambda_0
\end{eqnarray}
with
\begin{eqnarray}
\lambda_0= \frac{1}{2\mu_i \mu_j}(1-\mu_i^2 - \mu_j^2 +\lambda_{ij}^{1/2})
\ \ \ \ , \ \
\lambda_{1/2}= \frac{1}{2\mu_{j/i}} (1 \mp \mu_i^2 \pm \mu_j^2 -\lambda_{ij}^{
1/2})
\end{eqnarray}

\nn Note that $\Delta^R_{ij}$ has the same infrared divergence as $\Delta^V_{ij
}$ but with the opposite sign so that, as it should be, the sum of the two
contributions $\Delta_{ij}=\Delta^V_{ij}+\Delta^R_{ij}$ is infrared finite.

\bigskip

\nn {\bf 3.3 Equal squark mass case}

\bigskip

\nn In the case where the two scalar particles in the final state have equal
masses, the previous expressions simplify considerably. In terms of the
velocity of one of the final squarks one would have for the virtual and real
corrections
\begin{eqnarray}
\Delta_{ij}^V &=& -\log \mu^2_g \left[1+ \frac{1+\beta^2}{2\beta}
\log\frac{1-\beta}{1+\beta} \right] -2 +\log\frac{1-\beta^2}{4}
-\frac{1+\beta^2}{\beta} \log\frac{1-\beta}{1+\beta}
\non \\
& & + \frac{1+\beta^2}{\beta} \left[ \Li \left(\frac{1-\beta}{1+\beta}
\right) +\frac{1}{4} \log^2\frac{1-\beta}{1+\beta} + \log \beta \log
\frac{1-\beta}{1+\beta}+\frac{\pi^2}{3} \right]
\end{eqnarray}
\begin{eqnarray}
\Delta_{ij}^R &=& \log \mu^2_g \left[1+ \frac{1+\beta^2}{2\beta}
\log\frac{1-\beta}{1+\beta} \right] +\frac{3+7\beta^2}{2\beta^2}
+ \frac{3-6\beta^2-\beta^4}{4\beta^3} \log\frac{1-\beta}{1+\beta}
-4 \log \beta  \non \\
&&  +2 \log \frac{1-\beta^2}{4} + \frac{1+\beta^2}{\beta} \left[
3 \Li \left(\frac{1-\beta}{1+\beta} \right) + 2 \Li \left(-\frac{1-\beta}
{1+\beta} \right) -\frac{\pi^2}{3}  -\frac{5}{4} \log^2\frac{1-\beta}{1+\beta}
\right. \non \\
&& \left.  +\log\frac{1-\beta^2}{4\beta^2}\log\frac{1-\beta}{1+\beta}
+ \log\frac{2\beta}{1+\beta}\log\frac{1-\beta}{1+\beta}+2\log
\frac{4\beta}{(1+\beta)^2} \log\frac{1-\beta}{1+\beta} \right]
\end{eqnarray}
Adding these two contributions, one obtains the total gluonic QCD corrections
to squark pair production in the equal mass case
\begin{eqnarray}
\Delta_{ij} &=& \frac{3}{2} \frac{1+\beta^2}{\beta^2}+3\log\frac{1-\beta^2}{4}
-4\log \beta + \frac{3-10\beta^2-5\beta^4}{4\beta^3}\log\frac{1-\beta}{1+\beta}
+\frac{1+\beta^2}{\beta}
\non \\
& \times & \left[ 4\Li \left(\frac{1-\beta}{1+\beta}\right)
+2 \Li \left(-\frac{1-\beta}{1+\beta}\right)+2\log \beta\log
\frac{1-\beta}{1+\beta}- 3 \log\frac{1+\beta}{2}\log\frac{1-\beta}
{1+\beta} \right] \ \ \ \
\end{eqnarray}
This result is in agreement with the one obtained by Schwinger \cite{SCH} for
scalar QED, once a misprint in his eq.~(5.4--132) has been corrected; see also
Ref.~\cite{DH}. \s

Let us study the behaviour of this correction in the two limiting situations,
$\beta \ra 1$ or 0. In the limit where the squark mass is much smaller than the
center of mass energy, $\beta \ra 1$, the QCD correction approaches the value
$\Delta_{ij} \ra 3$, i.e. it is four times larger than the QCD corrections to
the production of massless spin--$\frac{1}{2}$ quarks. In the opposite limit
$\beta \ra 0$, i.e. for squark masses near the production threshold, one
obtains a correction $\Delta_{ij} \ra \pi^2/(2\beta)-2$, which exhibits the
well--known Coulomb singularity near threshold; in this case the perturbative
analysis is no longer reliable and one has to take into account
non--perturbative
effects as discussed in Ref.~\cite{BOUND}. \s

Finally, let us note that the expression eq.~(3.11) can be interpolated by the
Schwinger formula \cite{SCH}
\begin{eqnarray}
\Delta_{ij} \sim \frac{\pi^2}{2\beta}- \frac{1}{4}(1+\beta)(\pi^2-6)
\end{eqnarray}
which, up to an error of less than 2\%, reproduces the exact result given in
eq.~(3.11)

\renewcommand{\theequation}{4.\arabic{equation}}
\setcounter{equation}{0}

\subsection*{4. Gluino corrections}

Let us now discuss the supersymmetric corrections due to quark--gluino loops.
They consist of the vertex correction where the partner quark and a gluino are
exchanged, Fig.~1e, and of the contribution of the squark wave function
renormalization, Fig.~1f. In the general case where squark mixing is present,
leading to two final particles with different masses, the expressions of these
contributions are rather complicated.

\bigskip

\nn {\bf 4.1 Vertex Correction}

\bigskip

\nn The contribution of the quark--gluino vertex correction to the coupling of
the $Z$ boson to a pair of squarks $\tilde{q}_i \tilde{q}_j$ can be written as
\begin{eqnarray}
\Gamma^\mu_{ij} = \frac{-ie}{4c_Ws_W} \left[ a_{ij} k^\mu + \frac{4}{3}
\frac{\alpha_s}{\pi} \delta \Gamma^\mu_{ij} \right]
\end{eqnarray}
where
\begin{eqnarray}
\delta \Gamma^\mu_{ij} &=& v_q (\tilde{v}_i\tilde{v}_j+\tilde{a}_i\tilde{a}_j)
\left\{ q^\mu \left[ (2\tilde{m}_g^2+ 2m_q^2-\tilde{m}^2_{q_i}-\tilde{m}
^2_{q_j})C^+_{ij}+(\tilde{m}^2_{q_j}-\tilde{m}^2_{q_i})C^-_{ij} \right] \right.
\non \\
&+& \left. k^\mu \left[ (2\tilde{m}_g^2+ 2m_q^2+\tilde{m}^2_{q_i}+\tilde{m}^2_{
q_j})C^+_{ij}+(\tilde{m}^2_{q_i}-\tilde{m}^2_{q_j})C^-_{ij}
 +2 \tilde{m}_g^2 C^0_{ij} +B^0(s,m^2_q,m^2_q) \right]
\right\} \non \\
&+& a_q (\tilde{v}_i\tilde{a}_j+\tilde{a}_i\tilde{v}_j)
\left\{ q^\mu \left[ (2\tilde{m}_g^2- 2m_q^2-\tilde{m}^2_{q_i}-\tilde{m}^2_{
q_j})C^+_{ij}+(\tilde{m}^2_{q_j}-\tilde{m}^2_{q_i})C^-_{ij} \right] \right.
\non \\
&+& \left. k^\mu \left[ (2\tilde{m}_g^2- 2m_q^2+\tilde{m}^2_{q_i}+\tilde{m}^2_{
q_j})C^+_{ij}+(\tilde{m}^2_{q_i}-\tilde{m}^2_{q_j})C^-_{ij}
 +2 \tilde{m}_g^2 C^0_{ij} +B^0(s,m^2_q,m^2_q) \right]
\right\} \non \\
&+& 2\tilde{m}_g m_q q^\mu \left[2v_q(\tilde{v}_i\tilde{v}_j-\tilde{a}_i
\tilde{a}_j) C^-_{ij} - a_q (\tilde{v}_i\tilde{a}_j-\tilde{a}_i\tilde{v}_j)
C^0_{ij} \right] \non \\
&+& 2\tilde{m}_g m_q k^\mu v_q (\tilde{v}_i\tilde{v}_j-\tilde{a}_i\tilde{a}
_j) (2C^+_{ij} + C^0_{ij})
\end{eqnarray}
where, in terms of the scalar two--point and three point functions $B^0$ and
$C^0$ which are given in the Appendix, the functions $C^\pm_{ij} \equiv C^\pm
(\tilde{m}^2_{q_i},\tilde{m}^2_{q_j},s, m_q, m_q, \tilde{m}_g)$ are given by
\begin{eqnarray}
C^+_{ij}&=& -\frac{1}{2s\lambda_{ij}} \left\{ \frac{\tilde{m}^2_{q_i}-\tilde{m
}^2_{q_j}}{s} \left[ B^0(\tilde{m}^2_{q_i},\tilde{m}^2_{g},m_q^2)-
B^0(\tilde{m}^2_{q_j},\tilde{m}^2_{g},m_q^2)+ (\tilde{m}^2_{q_i}-\tilde{m}^2
_{q_j})C^0_{ij} \right] \right. \non \\
&& \hspace*{1.5cm} + B^0(\tilde{m}^2_{q_i},\tilde{m}^2_{g},m_q^2)+
B^0(\tilde{m}^2_{q_j},\tilde{m}^2_{g},m_q^2)-2B^0(s, m_q^2, m_q^2)
\non \\
&& \hspace*{1.5cm} \left. +(2m_q^2-2\tilde{m}_g^2-\tilde{m}^2_{q_i}-\tilde{m}
^2_{q_j}) C^0_{ij} \right\}
\end{eqnarray}
\begin{eqnarray}
C^-_{ij}&=& \frac{1}{2s\lambda_{ij}} \left\{\frac{2\tilde{m}^2_{q_i}+2\tilde{m
}^2_{q_j}-s}{s} \left[ B^0(\tilde{m}^2_{q_i},\tilde{m}^2_{g},m_q^2)- B^0(\tilde
{m}^2_{q_j},\tilde{m}^2_{g},m_q^2)+ (\tilde{m}^2_{q_i} - \tilde{m}^2_{q_j})
C^0_{ij} \right] \right. \non \\
&& \hspace*{1.5cm} + \frac{\tilde{m}^2_{q_i}-\tilde{m}^2_{q_j}}{s} \left[
B^0(\tilde{m}^2_{q_i}, \tilde{m}^2_{g},m_q^2)+
B^0(\tilde{m}^2_{q_j},\tilde{m}^2_{g},m_q^2)-2B^0(s, m_q^2, m_q^2)
\right. \non \\
&& \hspace*{1.5cm} \left. \left.
+(2m_q^2-2\tilde{m}_g^2-\tilde{m}^2_{q_i}-\tilde{m}^2_{q_j})
C^0_{ij} \right] \right\}
\end{eqnarray}

Note that while $C^+_{ij}$ is symmetric in the interchange of $\tilde{q_i}$ and
$\tilde{q_j}$, $C^-_{ij}$ is antisymmetric and vanishes in the equal mass case.
Note also that only the terms proportional to $k^\mu=k_1^\mu +k_2^\mu$ are
ultraviolet infinite [only the two--point function $B^0$ is divergent], this
has to be expected since the one--loop induced terms proportional to $q^\mu =
k_1^\mu-k_2^\mu$, which are absent at the tree--level, cannot be renormalized.
\s

The contribution to the coupling of the photon can be straightforwardly
derived from the previous expressions by setting to zero the terms proportional
to $a_q$; a further simplification is obtained by noting that $\tilde{v}_1
\tilde{v}_2+\tilde{a}_1\tilde{a}_2=0$ and discarding the terms proportional
to $q^\mu$ in the equal mass case [since we are interested in the interference
with the Born amplitude, these give rise to $k \cdot q=\tilde{m}^2_{q_1}-
\tilde{m}^2_{q_2}$ terms which vanish in this case]; one obtains
\begin{eqnarray}
\Gamma^\mu_{ij} = -ie e_q \left[ k^\mu \delta_{ij} +\frac{4}{3} \frac{\alpha_s}
{\pi} \delta \Gamma^\mu_{ij} \right]
\end{eqnarray}
with
\begin{eqnarray}
\delta \Gamma^\mu_{ij} &=& k^\mu (\tilde{v}_i \tilde{v}_j + \tilde{a}_i \tilde{
a}_j) \left[ (2\tilde{m}_g^2+ 2m_q^2+ \tilde{m}^2_{q_i}+\tilde{m}^2_{q_j} )C^+
_{ij}+2 \tilde{m}_g^2 C^0_{ij} + B^0(s,m^2_q,m^2_q)
\right] \delta_{ij} \non \\
&& + 2\tilde{m}_g m_q (\tilde{v}_i \tilde{v}_j - \tilde{a}_i \tilde{a}_j)
\left[k^\mu( 2 C^+_{ij}+ C^0_{ij}) + 2 q^\mu C^-_{ij} \right]
\end{eqnarray}
Note that for $i\neq j$, the contribution $\delta \Gamma_{ij}$ does not vanish
alone [because of the terms in the second line of the previous expression:
since the two masses $\tilde{m}_{q_i}$ and $\tilde{m}_{q_j}$ are not equal,
this leads to $k.q = \tilde {m}^2_{q_1}-\tilde{m}^2_{q_2} \neq 0$] and this
poses the problem of gauge invariance at the photon vertex; however, the
left--over piece in the amplitude
$q_\mu \delta \Gamma^\mu_{ij}$
\begin{eqnarray}
q_\mu \delta \Gamma^\mu_{ij} \sim 2 (k.q)C^+_{ij} + (k.q)C^0_{ij} +2
q^2C^-_{ij}
\ra B^0(\tilde{m}^2_{q_j}, \tilde{m}^2_{g},m_q^2) -
B^0(\tilde{m}^2_{q_i},\tilde{m}^2_{g},m_q^2)
\end{eqnarray}
is antisymmetric in the interchange of $i$ and $j$ so that the sum $q_\mu
\delta \Gamma^\mu_{12}+ q_\mu \delta \Gamma^\mu_{21}$, which enters in
the physical process, vanishes and the $\gamma \tilde{q_1}\tilde{q_2}$ is
indeed gauge invariant. This feature provides a good check of the calculation.
\s

\bigskip

\nn {\bf 4.2 Counterterms}

\bigskip

\nn One has then to include the renormalization of the squark wave function.
In the general case where squark mixing is allowed, this renormalization is a
bit more complicated than usual; this is because the wave--functions of the two
squarks are not decoupled. In addition one has also to renormalize the mixing
angle $\heta$. Taking into account the mixing, the renormalization of the
squark
wave functions and the mixing angle $\heta$ can be performed by making the
following substitutions in the Lagrangian eq.~(2.1)
\begin{eqnarray}
\tilde q_1 \rightarrow \left( 1+\delta Z_{11} \right)^{1/2} \tilde q_1+
\delta Z_{12} \tilde q_2 \ \ , \ \tilde q_2 \rightarrow \left(1+\delta Z_{22}
\right)^{1/2} \tilde q_2 + \delta Z_{21} \tilde q_1 \non \\
\heta \rightarrow \heta+\delta \heta \hspace*{5cm}
\end{eqnarray}
Under the substitution $\heta \ra \heta +\delta \heta$, the couplings $a_{ij}$
transform as $a_{ij} \ra a_{ij}+\delta a_{ij}$, with
\begin{eqnarray}
\delta a_{11}=2 a_{12} \delta \heta \ \ , \
\delta a_{22}=-2 a_{12} \delta \heta \ \ , \
\delta a_{12}=(a_{22}-a_{11}) \delta \heta
\end{eqnarray}
The full counterterms can then be included by shifting the couplings $a_{ij}$
in $\Gamma^{\mu}_{ij}$ of eq.~(4.1) and (4.5) by an amount
\begin{eqnarray}
a_{ij} \ra a_{ij}+\Delta a_{ij}
\end{eqnarray}
with
\begin{eqnarray}
\Delta a_{11} &=& \delta a_{11} + a_{11} \delta Z_{11} + 2 a_{12}
\delta Z_{21} \non \\
\Delta a_{22} & =&  \delta a_{22} + a_{22} \delta Z_{22} + 2 a_{12}
\delta Z_{12} \non \\
\Delta a_{12} &=& \delta a_{12} +\frac{1}{2}a_{12}(\delta Z_{11}+\delta Z_{22})
+ a_{11} \delta Z_{12} + a_{22} \delta Z_{21}
\end{eqnarray}
One has then to choose a renormalization condition which defines the mixing
angle $\heta$ [and hence $a_{12}$ through eq.~(4.9)]. We choose this condition
in such a way that the total quark--gluino QCD correction to the vertex $Z_\mu
\tilde q_1 \tilde q_2$ vanishes at zero--momentum transfer $q^2=0$; $\delta
a_{12}$ will be then given by
\begin{eqnarray}
\delta a_{12}=- \left[ \left. \frac{4}{3} \frac{\alpha_S}{\pi} \frac{1}{2}
\delta (\Gamma_{12}+\Gamma_{21}) \right|_{q^2=0} + \frac{1}{2} a_{12} (\delta
Z_{11}+ \delta Z_{22})\  +\ a_{11}\delta Z_{12}\ +\ a_{22} \delta Z_{21}
\right]
\end{eqnarray}
In the on--shell scheme, where the renormalized squark propagators are such
that their poles are at $p^2=\tilde{m}^2$ and the residues at the poles are
equal to unity, the counterterms $\delta Z_{ij}$ read
\begin{eqnarray}
 \delta Z_{11}= {\Sigma}_{11}'(m_{\tilde q_1}^2) \ \ , \
\   \delta Z_{22}={\Sigma}_{22}'(m_{\tilde q_2}^2)\ , \non \\
\delta Z_{12}=\frac{{\Sigma}_{12}(m_{\tilde q_2}^2)}{m_{\tilde q_2}^2-
m_{\tilde q_1}^2}\ ,   \  \delta Z_{21}=\frac{{\Sigma}_{21}(m_{\tilde q_1}^2)}
{m_{\tilde q_1}^2-m_{\tilde q_2}^2}
\end{eqnarray}
where $\Sigma_{ij}(m^2)$ and $\Sigma_{ii}'(m^2)= \partial \Sigma_{ii}(p^2)/
\partial p^2 |_{p^2=m^2}$ are given by
\begin{eqnarray}
\Sigma_{ij}(p^2) &= & \frac{4\alpha_S}{3 \pi} \left[ (\tilde{v}_i\tilde{v}_j
+\tilde{a}_i\tilde{a}_j) [A^0(m_{\tilde g}^2)+A^0(m_q^2)+ (m_{\tilde g}^2 +
m_q^2 -p^2) B_0(p^2, \tilde{m}_g^2,m_q^2)] \right.
\non \\
&& \left. \hspace*{1cm} + 2\tilde{m}_{g} m_q (\tilde{v}_i\tilde{v}_j-\tilde{a}
_i\tilde{a}_j) B^0(p^2, \tilde{m}_g^2,m_q^2) \right]
\end{eqnarray}
and
\begin{eqnarray}
\Sigma_{ii}^{'} (p^2) &=& - \frac{4\alpha_S}{3 \pi} \left[ (\tilde{v}_i^2+
\tilde{a}_i^2) [B_0(p^2, \tilde{m}_g^2,m_q^2) + (p^2-m_q^2-\tilde{m}_g^2)
B_0'(p^2, \tilde{m}_g^2,m_q^2)] \right. \non \\
& & \left. \hspace*{1cm} -2 m_q \tilde{m}_g (\tilde{v}_i^2- \tilde{a}_i^2)
B_0'(p^2, \tilde{m}_g^2,m_q^2) \right]
\end{eqnarray}

\nn with $A^0$ and $B^0$ the scalar one--point and two--point functions which
can be found in the Appendix.

\bigskip

\nn {\bf 4.3 Case of vanishing mixing angle}

\bigskip

\nn In the case where the mixing angle is set to zero, as is the case for the
SUSY partners of the light quarks and for stop and sbottom squarks when
$A_{t,b}=\mu=0$,
the situation simplifies considerably. One has just to consider separately
left-- and right--handed squarks, include the vertex correction [with equal
squark masses] and the squark wave--function renormalization [with the ones of
$\tilde{ q}_1$ and $\tilde{q}_2$ now decoupled]. The sum of these two
contributions is finite and no mixing angle renormalization is needed. \s

The coupling of the photon and the Z boson to a pair of squarks including the
quark--gluino vertex correction and the wave--function renormalization, will
read at ${\cal O}(\alpha_S)$
\begin{eqnarray}
\Gamma^\mu_{\gamma \tilde{q}_i\tilde{q}_i} &=& -ie e_q (k_1+k_2)^\mu \left(1 +
\frac{4}{3} \frac{\alpha_S}{\pi} \frac{1}{2} \tilde{\Delta}_{ii}^\gamma \right)
\non \\
\Gamma^\mu_{Z \tilde{q}_i\tilde{q}_i} &=& -\frac{ie}{4c_Ws_W} (v_q \pm a_q)
(k_1+k_2)^\mu \left(1 + \frac{4}{3} \frac{\alpha_S}{\pi} \frac{1}{2}
\tilde{\Delta}_{ii}^Z \right)
\end{eqnarray}
where in the case of the $Z$ boson, the upper sign is for $i=1=$L and the
lower sign for $i=2=$R; the corrections factor $\tilde{\Delta}_{ii}^Z$ is
given by
\begin{eqnarray}
\tilde{\Delta}_{ii}^Z &=& 2 \left[ \tilde{m}_g^2+ \tilde{m}^2_{q} + \frac{v_q
\mp a_q}{v_q \pm a_q} m_q^2 \right]C_+ + 2 \tilde{m}_g^2 C_0 +
B_0(s,m^2_{q},m^2_{q}) \non \\
&&  - B_0(\tilde{m}^2_{q},\tilde{m}^2_{g}, m_q^2)
- (\tilde{m}^2_{q}-m_q^2-\tilde{m}^2_{g}) B_0'
(\tilde{m}^2_{q},\tilde{m}^2_{g},m_q^2)
\end{eqnarray}
For the correction to the photon vertex, one just has to replace $v_q$ by $e_q$
and set $a_q=0$ in the previous equation; one obtains
\begin{eqnarray}
\tilde{\Delta}_{ii}^\gamma &=& \tilde{\Delta}_{ii} =2(\tilde{m}_g^2+
\tilde{m}^2
_{q} + m_q^2) C_+ + 2\tilde{m}_g^2 C_0 + B_0(s,m^2_{q},m^2_{q}) \non \\
&&  - B_0(\tilde{m}^2_{q},\tilde{m}^2_{g}, m_q^2)
- (\tilde{m}^2_{q}-m_q^2-\tilde{m}^2_{g}) B_0'
(\tilde{m}^2_{q},\tilde{m}^2_{g},m_q^2)
\end{eqnarray}
In terms of the two and three point scalar functions $B_0$ and $C_0$, the
function $C_+$ in the equal mass case simplifies to
\begin{eqnarray}
C_+ &=& -\frac{1}{s\beta^2_q} \left[ B_0(\tilde{m}^2_{q},\tilde{m}^2_{g},m_q^2)
-B_0(s, m^2_q, m^2_q)+(m_q^2-\tilde{m}_g^2-\tilde{m}^2_{
q})C_0 \right]
\end{eqnarray}
Note that the correction to the $Z$--squarks vertex can be written as
\begin{eqnarray}
\tilde{\Delta}_{ii}^{Z} = \tilde{\Delta}_{ii} + \tilde{\delta}_{ii} \ \ \ \ ,
\ \ \ \delta_{ii} = \mp \frac{4a_q/v_q}{1\pm a_q/v_q} m_q^2 C_+
\end{eqnarray}
which exhibits the fact that for the superpartners of the light quarks, where
one can set $m_q=0$, one would have $\tilde{\Delta}_{ii}^{Z}=\tilde{\Delta}_
{ii}^{\gamma}=\tilde{\Delta}_{ii}$.

\newpage

\renewcommand{\theequation}{5.\arabic{equation}}
\setcounter{equation}{0}

\subsection*{5. Numerical results and discussions}

In this section we will discuss the magnitude of these QCD corrections,
restricting ourselves to the case of vanishing mixing angle [and hence,
to the case of degenerate final state squarks] for which the numerical
analysis is
simpler. \s

First of all, Fig.~2 shows the magnitude of the QCD correction factor
$\Delta_{ii}$ due to virtual gluon exchange and real gluon emission, as a
function of the velocity of the final squarks, $\beta_q$. As previously
mentioned, $\Delta_{ii}$ rapidly increases from the value $\Delta_{ii}=$ 3 for
massless squarks (i.e. $\beta_q =1)$ to reach an infinite value near the
production threshold ($\beta_q=0)$. Close to the latter value, perturbation
theory is no more reliable and non--perturbative effects have to be taken into
account. \s

In Fig.~3 and Fig.~4, we display the QCD correction factor
$\tilde{\Delta}_{ii}$ due to quark-gluino exchange as a function of the squark
masses and for two center of mass energy values: $\sqrt{s}=M_Z$ (Fig.~3) and
$\sqrt{s}=500$ GeV (Fig.~4). While for the SUSY partners of light quarks, only
one type of correction $\tilde{\Delta}_{ii}$ (in Figs.~3a and 4a) is present,
three types of corrections are needed for the scalar partners of the top quark:
$\tilde{\Delta}_{ii}^\gamma$ (in Figs.~3b and 4b) for photon exchange and
$\tilde{\Delta}_{ii}^Z$ for $Z$ boson exchange and for both the left--handed
and
right--handed squarks (in Figs.~3c,4c and Fig.~3d,4d respectively). \s

As representative values of the gluino mass, we have chosen $\tilde{m}_g=5$
GeV, 100 GeV and 250 GeV for LEP energies and $\tilde{m}_g=5$ GeV, 100 GeV and
500 GeV at $\sqrt{s}=500$ GeV. We have allowed for the possibility that the
gluinos might be very light, $\tilde{m}_g \sim 5$ GeV. Indeed, present
experimental data do not exclude this scenario and a recent analysis
\cite{Farrar} shows that gluinos with masses in the 3--5 GeV range are still
allowed\footnote{Note that, even this window would be sufficient to allow for a
substantial modification of the running of the strong coupling constant
\cite{Kuhn} which is the original motivation for the renewed interest in this
scenario.}. In this case, the bounds on squark masses derived from searches for
events with a large missing transverse momentum at hadron colliders \cite{CDF}
might be invalidated\footnote{This is because of the fact that squarks will
dominantly decay into light gluinos, which loose a very large fraction of their
energy in QCD radiation before they decay \cite{Farrar}, therefore leading to a
soft missing transverse momentum spectrum.}, and thus, the only valid bounds on
these masses would come from the negative searches of squark pairs in $Z$
decays [7]. These bounds,
in fact, will possibly be altered by the inclusion of the
QCD corrections that we computed here, and this motivates our discussion of
these contributions for LEP100 energies. \s

At LEP100, the QCD correction factors due to the contributions of quark--gluino
loops are displayed in Fig.~3a to 3d for three values of $\tilde{m}_g=5$ GeV,
100 GeV and 250 GeV. They are practically constant and rather small,
$|\tilde{\Delta}_{ii}^V |<1$, except for the scalar partners of light quarks
and for $\tilde{m}_g=5$ GeV, in which case
$\tilde{\Delta}_{ii}^\gamma$ varies from $-4$ for $\tilde{m}_q \sim 10$
GeV to $-1.2$  near the production threshold. The
reason is that for large internal particle masses, as is the case at this
energy for the other two values of $\tilde{m}_g$ and for the top quark for
which we took a mass of $m_t=175$ GeV \cite{TOP}, the correction is
proportional to the inverse of the mass squared; this ensure the proper
decoupling of the amplitude in this limit. \s

For the production of squark pairs at a future high--energy $\ee$ collider with
a center of mass energy of $\sqrt{s}=500$ GeV, the QCD correction factors are
displayed in Fig.~4a to 4d for three values of $\tilde{m}_g=5$ GeV, 100 GeV and
500 GeV. These factors are larger than in the previous case, especially for the
scalar partners of top quarks. This is due to the fact that the top mass [as
well as a gluino mass of 100 GeV] is no more much larger than the beam energy
and the contributions do not decouple yet. However, the corrections are still
relatively small except in the light gluino window, and in some cases for
gluino masses of the order of 100 GeV. Note that the dips in some of the curves
correspond to the opening of the $\tilde{q} \ra q+\tilde{g}$ channel. \s

The QCD corrections to the cross section can be readily obtained by replacing
in eq.~(2.9), the bare couplings $e_q$ and $v_{\tilde{q}_i}$ with their
renormalized values of eqs.~(4.16--4.18) [to include the corrections
originating from quark--gluino loops] and multiplying the bare cross section
by a factor $(1+4\alpha_S/(3\pi)\Delta_{ii})$ as in eq.~(3.1) [to include
the standard gluonic corrections]. Writing as in eq.~(4.20), $\tilde{\Delta}_{
ii}^{Z} = \tilde{\Delta}_{ii} + \tilde{\delta}_{ii}$, one obtains for the
QCD corrected cross section normalized to the Born term
\begin{eqnarray}
\Delta \sigma \equiv
\frac{ \sigma (e^+ e^- \rightarrow \tilde{q}_i \tilde{q}_i)}{\sigma^B (e^+ e^-
\rightarrow \tilde{q}_i \tilde{q}_i)}-1= \frac{4}{3}
\frac{\alpha_S}{\pi} \left[ \Delta_{ii} + \tilde{\Delta}_{ii} + \frac{1}{2}
\tilde{\delta}_{ii} \delta\sigma_{ii} \right]
\end{eqnarray}
with
\begin{eqnarray}
\delta \sigma_{ii} &=& \left[- \frac{e_q v_e v_{q_i} }{16c_W^2s_W^2}
\frac{s}{s-M_Z^2}+\frac{(a_e^2+v_e^2)v_{q_i}^2}{128s_W^4c_W^4}\frac{s^2}{(s-
M_Z^2)^2} \right] \non \\
&& \times \left[ e_q^2 - \frac{e_q v_e v_{q_i} }{16c_W^2s_W^2}
\frac{s}{s-M_Z^2}+\frac{(a_e^2+v_e^2)v_{q_i}^2}{256s_W^4c_W^4}\frac{s^2}{(s-
M_Z^2)^2} \right]^{-1}
\end{eqnarray}
On the $Z$ resonance, one can neglect the photon exchange and the
interference between $\gamma$ and $Z$ exchanges, and the expression of
$\Delta \sigma$ simplifies to
\begin{eqnarray}
\Delta \sigma \equiv
\frac{ \sigma (e^+ e^- \ra Z \rightarrow \tilde{q}_i \tilde{q}_i)}{\sigma^B
(e^+ e^- \ra Z \rightarrow \tilde{q}_i \tilde{q}_i)}-1= \frac{4}{3}
\frac{\alpha_S}{\pi} (\Delta_{ii} +\tilde{\Delta}_{ii}^Z )
\end{eqnarray}
In the case of the SUSY partners of light quarks, $\tilde{\delta_{ii}}=0$
and, independently of the center of mass energy, one simply has
\begin{eqnarray}
\Delta \sigma \equiv
\frac{ \sigma (e^+ e^- \rightarrow \tilde{q}_i \tilde{q}_i)}{\sigma^B (e^+ e^-
\rightarrow \tilde{q}_i \tilde{q}_i)}-1= \frac{4}{3} \frac{\alpha_S}{\pi}
(\Delta_{ii} + \tilde{\Delta}_{ii})
\end{eqnarray}

The deviation from unity of the QCD corrected cross section normalized to the
Born cross section as a function of the squark masses and for the previous
choices of the gluino masses is displayed in Fig.~5 and Fig.~6 for center
of mass energies of respectively $\sqrt{s}=M_Z$ and $\sqrt{s}=500$ GeV.
In these figures, we have used for the strong coupling constant the value
$\alpha_S (M_Z^2)\simeq 0.12$ as determined from various measurements in $Z$
decays \cite{Alphas}; at an energy of 500 GeV, the coupling will run down to
$\alpha_S \sim 0.11$.

\bigskip

As one might expect from the previous discussion, the largest part of the QCD
corrections is due to the standard gluonic corrections. The effect of the
gluino-quark loops at LEP energies is important only in the case of the
partners of light quarks when the gluino is very light, especially for small
squark masses where the correction can even flip sign. At 500 GeV, the effect
of gluino masses of the order of 100 GeV is also significant and in general,
reduces the size of the QCD correction. \s

For large values of the gluino mass, the quark--gluino contribution decouples
and only the ``standard" gluonic contributions are left. In this case, even for
squark masses very small compared to the center of mass energy of the collider,
the QCD corrections enhance the cross section by more than $15\%$. The
correction increases with the squark mass, and already for $\beta_q=0.5$, one
has a correction of more than $50\%$. These corrections are therefore much
larger than the ones affecting the pair production in $\ee$ collisions of
spin 1/2 particles with the same mass \cite{QCD}. \s

\subsection*{6. Summary}

In this paper, we have calculated the QCD radiative corrections to the
production of the supersymmetric partners of quarks in $\ee$ collisions. We
have taken into account the mixing between left-- and right--handed scalar
quarks, and this led us to treat the case where the two final state particles
have different masses. We included both the standard gluonic corrections
consisting of virtual gluon exchange and real gluon emission (which, in
the limiting case where the scalar particles are degenerate in mass, have been
calculated long time ago by Schwinger for scalar QED) and also the
genuine SUSY QCD corrections due to quark--gluino loops (which do not appear in
scalar QED). For both types of corrections, complete analytical expressions
were given in the general case. \s

We have then discussed the magnitude of these QCD corrections in the case of
vanishing mixing angle. We have shown that the corrections due to quark--gluino
loops are important only when the gluinos have relatively small masses, and
in this case, particularly for the production of the partners of light quarks.
For large gluino masses, these corrections decouple as they should, and
only the contributions due to gluon exchange and real gluon emission are left
in this limit. The latter corrections are very important since, even for
massless squarks where they are minimal, they enhance the cross section by
more than $15\%$, four times as much as the QCD corrections affecting the
production of spin 1/2 massless quarks. \s

\vspace*{1cm}

\nn {\bf Acknowledgements:} We thank M. Drees and P. Ouellette for discussions.

\newpage

\renewcommand{\theequation}{A.\arabic{equation}}
\setcounter{equation}{0}

\subsection*{Appendix: Scalar Loop Integrals}

\nn In this Appendix we collect expressions that allow to evaluate the
loop functions that appear in section 4.
The scalar one, two and three point functions, $A_0, B_0$ and $C_0$
are defined as
\begin{eqnarray}
A_0(m_0) &=& \frac{(2\pi\mu)^{n-4}}{i\pi^2} \int \ \frac{d^nk}{k^2-m_0^2+
i\epsilon} \non \\
B_0(s,m_1,m_2) &=& \frac{(2\pi\mu)^{n-4}}{i\pi^2} \ \int \frac{d^nk}{
(k^2-m_1^2 + i\epsilon) [(k-q)^2-m_2^2+i \epsilon] } \\
C_0( m_1, m_2, m_3) &=& \frac{(2\pi\mu)^{n-4}}{i\pi^2} \int \frac{d^nk}{
[(k-p_1)^2-m_1^2 +i\epsilon][(k-p_2)^2-m_2^2 +i\epsilon](k^2-m_3^2+i\epsilon)}
\non
\end{eqnarray}

\nn Here $n$ is the space--time dimension and $\mu$ the renormalisation scale.
After integration over the internal momentum $k$, the function $A_0$ is
given by [$\gamma_E$ is Euler's constant]:
\begin{eqnarray}
A_0(m_0)= m^2_0 \left[ 1+ \Delta_0 \right]  \ \ , \hspace*{1cm}
\Delta_i = \frac{2}{4-n} - \gamma_E + \log (4 \pi) +\log \frac{\mu^2}{m_i^2}
\end{eqnarray}
\nn  The function $B_0$ and its
derivative with respect to $s$, $B'_0$, are given by
\begin{eqnarray}
B_0(s,m_1,m_2) &=& \frac{1}{2}(\Delta_{1}+ \Delta_{2}) +2 +\frac{m_1^2-m_2^2}
{2s} \log \frac{m_2^2}{m_1^2} +\frac{x_+ -x_-}{4s} \log \frac{x_-}{x_+} \non \\
B_0'(s,m_1,m_2) &=& -\frac{1}{2s} \left[ 2+\frac{m_2^2-m_1^2}{s} \log
\frac{m_1^2} {m_2^2} + \frac{2}{s} \frac{(m_1^2-m_2^2)^2 - s(m_1^2+m_2^2)}
{x_+-x_-} \log\frac{x_-}{x_+} \right]
\end{eqnarray}
\nn with
\begin{eqnarray} \label{a4}
x_\pm =s- m_1^2 -m_2^2 \pm \sqrt{ s^2-2s (m_1^2+m_2^2)+(m_1^2 - m_2^2 )^2}
\end{eqnarray}
\nn Note that the $x_\pm$ can be complex; however we have ignored the imaginary
parts of $B_0$ and $B_0'$ since they are not relevant for us: to
next--to--leading order we are only interested in the interference between the
(real) tree--level and one--loop amplitudes.

\vspace*{5mm}
\nn We also need the three point scalar function $C_0$ which can be written in
integral form as
\begin{eqnarray} \label{a5}
C_0(\tilde{m}^2_{1}, \tilde{m}_2^2,s,m_1,m_2,m_3) = - \int_0^1 dy \int_0^y dx
\left[ ay^2+bx^2+cxy +dy+ex+f \right]^{-1}
\end{eqnarray}
where
\begin{eqnarray} \label{a6}
a=m_q^2 \ , \ \ b=s \ , \ \ c=-s \ , \ \ d = m_2^2 -m_3^2 - m_q^2 \ ,
\ \ e = m_1^2 -m_2^2 \ , \ \ f=m_3^2 -i \epsilon \hspace*{3mm}
\end{eqnarray}
\nn The full analytical expression of $C_0$ in terms of Spence functions can be
found in Ref.~\cite{TV}.

\newpage

\subsection*{Figure Captions}

\vspace*{0.5cm}

\renewcommand{\labelenumi}{Fig. \arabic{enumi}}
\begin{enumerate}
\item  
Feynman diagrams for the QCD corrections to the decay of a gauge boson into
squark pairs: tree--level diagram (1a);  vertex (1b) and self--energy (1c)
standard gluonic corrections and real gluon emission (1d) ; vertex (1e) and
self--energy (1f) corrections due to quark--gluino loops.

\vspace{5mm}
\item  
The QCD correction factor $\Delta_{ii}$ due to virtual gluon exchange and real
gluon emission as a function of the velocity of the final squarks $\beta$. The
two squarks are asummed to be degerate in mass.

\vspace{5mm}
\item  
The QCD correction factors due to virtual quark--gluino exchange as a function
of the squark mass and for selected values of the gluino mass at $\sqrt{s}=
M_Z$: $\tilde{\Delta}_{ii}$ for the partners of light quarks (3a),
$\tilde{\Delta}_{ii}^\gamma$ for degenerate stop squarks (3b) and
$\tilde{\Delta}_{ii}^Z$ for the left--handed (3c) and right--handed stop
squarks (3d). The full line is for $\tilde{m}_g=250$ GeV, the dashed line
for $\tilde{m}_g=100$ GeV and the dot--dashed line for $\tilde{m}_g=5$ GeV.

\vspace{5mm}
\item  
The QCD correction factors due to virtual quark--gluino exchange as a function
of the squark mass and for selected values of the gluino mass at $\sqrt{s}=
500$ GeV: $\tilde{\Delta}_{ii}$ for the partners of light quarks (4a),
$\tilde{\Delta}_{ii}^\gamma$ for degenerate stop squarks (4b) and
$\tilde{\Delta}_{ii}^Z$ for the left--handed (4c) and right--handed stop
squarks (4d). The full line is for $\tilde{m}_g=500$ GeV, the dashed line is
for $\tilde{m}_g=100$ GeV and the dot--dashed line for $\tilde{m}_g=5$ GeV.

\vspace{5mm}
\item  
The deviation from unity of the fully QCD corrected cross section normalized to
the Born cross section as a function of the squark masses and for selected
values of the gluino mass at $\sqrt{s}=M_Z$: for the partners of light quarks
(5a), and for the left--handed (5b) and right--handed stop squarks (5c).
The full line is for $\tilde{m}_g=250$ GeV, the dashed line
for $\tilde{m}_g=100$ GeV and the dot--dashed line for $\tilde{m}_g=5$ GeV.

\vspace{5mm}
\item  
The deviation from unity of the fully QCD corrected cross section normalized to
the Born cross section as a function of the squark masses and for selected
values of the gluino mass at $\sqrt{s}=500$ GeV: for the partners
od light quarks
(6a), and for the left--handed (6b) and right--handed stop squarks (6c).
The full line is for $\tilde{m}_g=500$ GeV, the dashed line is
for $\tilde{m}_g=100$ GeV and the dot--dashed line for $\tilde{m}_g=5$ GeV.

\end{enumerate}

\end{document}